\documentclass[pre, 11pt, showkeys, superscriptaddress, amsmath, amssymb, floatfix]{revtex4-1} 

\usepackage{graphicx}  
\usepackage{graphics}
\usepackage{dcolumn}   
\usepackage{bm}        
\usepackage{amssymb}   
\usepackage{color}
\usepackage{amsmath}

\definecolor{darkred}{RGB}{139,0,0}
\definecolor{chartreuse}{RGB}{127,255,0}
\definecolor{goldenrod}{RGB}{218,165,32}
\definecolor{gray}{RGB}{127,127,127}

\definecolor{Magenta}{RGB}{255, 0,255}
\definecolor{Orange}{RGB}{255,165, 0}
\definecolor{Gray}{RGB}{127,127,127}

\begin{document}



\title{Quantifying patterns of research interest evolution}
\author{Tao Jia}
\email{tjia@swu.edu.cn}
\affiliation{College of Computer and Information Science, Southwest University, Chongqing, 400715, P. R. China}
\affiliation{Laboratory for Software and Knowledge Engineering, Southwest University, Chongqing, 400715, P. R. China}
\author{Dashun Wang}
\email{dashun.wang@northwestern.edu}
\affiliation{Kellogg School of Management, Northwestern University, Evanston, IL 60208, USA.}
\affiliation{Northwestern Institute on Complex Systems (NICO), Northwestern University, Evanston, IL 60208, USA}
\affiliation{McCormick School of Enginnering and Applied Sciences, Northwestern University, Evanston, IL 60208, USA}
\author{Boleslaw K. Szymanski}
\email{szymab@rpi.edu}
\affiliation{Social Cognitive Networks Academic Research Center, Rensselaer Polytechnic Institute, Troy, NY, 12180 USA}
\affiliation{Department of Computer Science, Rensselaer Polytechnic Institute, Troy, NY, 12180 USA}
\affiliation{Spo\l{}eczna Akademia Nauk, 90-113 \L{}\'{o}d\'{z}, Poland}

\maketitle 


{\bf Our quantitative understanding of how scientists choose and shift their research focus over time is highly consequential, because it affects the ways in which scientists are trained, science is funded, knowledge is organized and discovered, and excellence is recognized and rewarded \cite{kuhn2012structure,de1986little,merton1973sociology,rzhetsky2015choosing, kuhn1979essential, sinatra2016quantifying, wang2013quantifying, foster2013tradition, clauset2015systematic}. Despite extensive investigations of various factors that influence a scientist's choice of research topics \cite{jones2011age, packalen2015multiple,duch2012possible, west2013role,clauset2015systematic, malmgren2010role,guimera2005team,jones2008multi,hoonlor2013trends,mcnally2011discovery,ericsson2006influence,foster2013tradition, azoulay2011incentives, bergstrom2016scientists}, quantitative assessments of mechanisms that give rise to macroscopic patterns characterizing research interest evolution of individual scientists remain limited. Here we perform a large-scale analysis of publication records, finding that research interest change follows a reproducible pattern characterized by an exponential distribution. We identify three fundamental features responsible for the observed exponential distribution, which arise from a subtle interplay between exploitation and exploration in research interest evolution \cite{kuhn1979essential, march1991exploration}. We develop a random walk based model, allowing us to accurately reproduce the empirical observations. This work presents a quantitative analysis of macroscopic patterns governing research interest change, discovering a high degree of regularity underlying scientific research and individual careers.}

``The essential tension" hypothesis set forth by Thomas Kuhn \cite{kuhn1979essential} has vividly highlighted the conflicting demands of scientific careers that require both exploration and exploitation \cite{march1991exploration,rzhetsky2015choosing,foster2013tradition}. Indeed, career advancement, from promotion to obtaining grants, demands a steady stream of publications, which is often achieved through uninterrupted yet incremental contributions to existing, established research agenda. In contrast, frequent changes in research topics invite risk of failure and decreased productivity. The disciplinary boundaries, arising from such factors as implicit culture, tacit and accumulated knowledge \cite{vilhena2014finding, hidalgo2015information} and peer recognition \cite{merton1973sociology,de1966collaboration}, together with intensifying specialization in science and engineering disciplines \cite{jones2009burden}, make radical shifts, such as moving from chemical biology to high energy physics, extremely unlikely, if at all possible. On the other hand, although steady and focused research portfolio helps scientists stay productive, it potentially undermines chances for originality \cite{foster2013tradition}. Indeed, innovative and novel insights often emerge from encountering new challenges and opportunities associated with venturing into new topics and/or incorporating them into existing research agenda \cite{rzhetsky2015choosing,uzzi2013atypical,azoulay2011incentives,youn2015invention,guimera2005team}.

Given the broad impact on individual careers and strong implications for science and innovation policy, there is an urgent need for quantitative approaches to understanding the nature of research interest change undertaken by individual scientists throughout their careers. This becomes ever more so with the accelerating scale and complexity of scientific enterprise \cite{de1986little, cokol2005emergent, jones2009burden,sinatra2015century}. A variety of microscopic factors have been identified that drive a scientist's choice of research problems, ranging from age \cite{jones2011age, packalen2015multiple} to gender \cite{duch2012possible, west2013role}, to training and mentorship \cite{clauset2015systematic, malmgren2010role}, from funding or collaboration opportunities \cite{guimera2005team,jones2008multi,hoonlor2013trends}, to serendipity \cite{mcnally2011discovery}, to scientist's attitudes and abilities \cite{ericsson2006influence}, including risk aversion and creativity \cite{foster2013tradition, azoulay2011incentives, bergstrom2016scientists}. Yet, little is known about the macroscopic patterns underlying the research interest evolution. Recent advances in complex systems have uncovered regularities in various dynamical processes, which give rise to a family of powerful yet flexible statistical models describing processes as diverse as human mobility \cite{gonzalez2008understanding, song2010modelling, simini2012universal, yan2014universal}, temporal dynamics \cite{barabasi2005origin,malmgren2008poissonian,zhao2013emergence} and the evolution of complex networks \cite{Barabasi-Science-99,scholtes2014causality,holme2012temporal}. This prompts us to ask: To what degree could the research interest evolution be captured by a simple model?

Here we aim to systematically address this question by first identifying patterns in the working scientists' research agendas as their careers progress. Using articles published by American Physical Society (APS) journals covering over 30 years (1976 - 2009) \cite{zhang2013characterizing,radicchi2009diffusion} and through a careful and extensive author name disambiguation process \cite{deville2014career, sinatra2016quantifying}, we collect individual authors' publication records over time (Supplementary Note 1). We further take advantage of the Physics and Astronomy Classification Scheme (PACS) codes used by APS to classify topics in physics. Indeed, among all identifiers for research topics, PACS codes stand out in its frequency of use \cite{herrera2010mapping,radicchi2011rescaling, pan2012evolution,wei2013scientists,sinatra2015century,shen2016interrelations}. This is partly because unlike topics defined by keywords, which are often created in an ad-hoc, unstructured manner, the PACS code classification relies on both crowd wisdom of working scientists and expert opinions of journal editors, offering a systematic representation of a paper's subject. There are 67 main topics defined by the first two digits of the PACS codes, covering diverse topics ranging from general relativity and gravitation to nuclear structure to superconductivity. By sorting these PACS codes, we obtain a {\it topic tuple} for each paper, representing its research subject as a combination of topics the paper studies (Fig. 1) \cite{youn2015invention,sinatra2015century}. For a given set of papers published by a scientist, we generate a {\it topic vector}, whose element measures the weighted occurrence of each topic (see Methods). Hence this vector captures not only the collection of topics a scientist studied but also the level of involvements in each of these topics. Consequently, this vector represents a multi-dimensional measure of research interest revealed in the series of published papers.

We compose two topic vectors based on the first and last $m$ papers of the scientist ($g_i$ and $g_f$ respectively), capturing the research interest at the earliest and the latest stages of the career (Fig. 1). Using the complementary cosine similarity between $g_i$ and $g_f$, we quantify the interest change $J$ of a scientist along the career as: 
\begin{equation}
J = 1 - \frac{g_i \cdot g_f}{\|g_i\| \|g_f\|}.\label{eq:J}
\end{equation}
Eq. (\ref{eq:J}) captures research interest change resulting from change of topics or from change of engagement in topics, providing an effective quantification on the extent of change. $J = 0$ indicates that two topic vectors $g_i$ and $g_f$ are identical, capturing the fact that the author not only studied the same set of topics at the two stages of the career but also was involved in each of these topics with the same weight. $J=1$ corresponds to a complete interest change in which a researcher does not engage in any initial topic of interest. We choose $m=8$. As a result, our analyses are based on 14,715 scientists who authored at least $2m=16$ papers included in our dataset. We report analyses based on other $m$ values ($m=6$ and $m=10$, Supplementary Note 2) and find that our results are insensitive to the choice of $m$. To take into account other factors that can play a role in quantifying research interest change, we further perform additional three measurements. First, to avoid the gaps between the two sets of papers, we take $2m$ consecutive papers starting at a randomly chosen paper and measure the interest change based on the two adjacent $m$ paper sequences (Supplementary Note 3). Furthermore, to eliminate the effects of different publication rates, we measure interest change $J$ within scientists who publish at similar rates (Supplementary Note 4). Finally, as research interest is associated with time, we measure interest change based on two sets of papers published over the same time period at the early and late stage of a career (Supplementary Note 5). We obtain similar observations in all these three measurements, demonstrating that the discovered patterns in research interest evolution do not rely on a specific metric chosen.

The quantification of research interest change allows us to measure its distribution within the population. We find that the fraction of scientists $P$ decays with the extent of interest change $J$ (Figs. 2a), which can be well fitted with an exponential function (see Supplementary Note 6 for fitting statistics). This exponential decay indicates that most scientists are characterized by little change in their research interests and the probability of making a leap decreases exponentially with its range. At the same time, the fact that histogram $P(J)$ is positive in the full range of domain $[0, 1]$ indicates that large changes in a career, such as switching to completely different areas, do occur, albeit very rarely. These observations raise an interesting question: What forms of interest change distribution should we expect? To answer this question, we uncover three features characterizing research interest evolution. To illustrate how they shape the distribution of interest change, we perform three ``experiments'' in which the interest change distribution is re-measured using modified publication sequences of individual scientists, which further demonstrates the non-trivial nature of the exponential distribution observed.

{\it Heterogeneity}: The frequencies of topic tuple occurrences in a scientist's publication sequence follow a power-law distribution (Fig. 2b), demonstrating the heterogeneity in individual's engagement in different subjects. Indeed, a scientist's research agenda contains core research subjects that are repeatedly investigated coupled with other more peripheral ones that may only be touched upon occasionally. To examine the effect of heterogeneity on the interest change, we modify each publication sequence by retaining only the first occurrence of each topic tuple and removing the subsequent ones (see Supplementary Fig. S1a for the illustration of this procedure and Supplementary Note 7 for more details). The interest change distribution measured over the modified sequences reaches a peak at an intermediate value followed by a gradual decay (Fig. 2c), in sharp contrast to the monotonic, exponential decrease observed in the real data.

{\it Recency}: Denoting the number of papers between two consecutive uses of a topic tuple by $\Delta n$ (see Methods), we compare the distribution $P(\Delta n)$ with that of the reshuffled publication sequence $P_\text{o}(\Delta n)$. We find that the ratio $P(\Delta n) / P_\text{o}(\Delta n)$ decreases with $\Delta n$, going from above to below 1 (Fig. 2d). This implies that scientists are more likely to publish on subjects recently studied in real sequences than in reshuffled sequences. To further explore this feature, we measure the relationship between the probability that a distinct topic tuple is re-studied ($\Pi$) and the order of its first occurrence in a scientist's publication sequence (see Methods). As shown in Fig. 2e, the relationship obtained clearly demonstrates that scientists are more likely to publish on subjects studied recently than on those investigated long time ago. The recency feature suggests that scientists tend to avoid going back to the original research subjects once they have moved to other ones, which consequently drives them to explore new research subjects. This is affirmed in our experiment where the interest change distribution is measured over the reshuffled publication sequences (see Supplementary Fig. S1b for the illustration of this procedure). The obtained distribution remains to be exponential, but with much steeper drop than the original distribution, which makes the large extent of interest change significantly less frequent than what is actually observed (Fig. 2f).

{\it Subject proximity}: Knowledge is characterized by underlying topical geometry that imposes varying inherent distances between pairs of research subjects \cite{jones2009burden, shi2015weaving, boyack2005mapping}. When a scientist changes the research subject, she is more likely to choose a subject related to the current one than moving to a totally new field, implying that research interest change is affected by subject proximity. To verify this insight, we modify each publication sequence by replacing distinct topic tuples in one's publications with ones randomly drawn (see Supplementary Fig. S1c for the illustration of this procedure). The resulting interest change distribution shows most scientists having a large extent of change (Fig. 2g), demonstrating the effect of subject proximity on research interest evolution.

The three features reveal rich insights offered by the observed distribution of research interest change. Both heterogeneity and subject proximity arise from exploitation of the current field that help stabilize the research interest. Lacking either of the two, $P$ would have been characterized by a distribution with a larger mean. Yet, the recency feature resulting from the exploration of new areas destabilizes research interest. If it were not for the exploitation, the interest change would have been much more limited given the heterogeneity and subject proximity features. Together they provide us with empirical basis to build a statistical model for individual careers with varying research subjects over time.

Here we consider scientific research as a random walk, following Sir Isaac Newton's retrospection that during his scientific career he was like ``{\it \ldots a boy playing on the seashore \ldots finding \ldots a prettier shell than ordinary}" \cite{mandelbrote2001footprints}. Such a ``seashore" is represented by a 1-D lattice in our model and a ``shell" corresponds to a scientific finding that yields a paper (Fig. 3a). The locations at which shells can be found are located with certain probability $p$ at the sites of lattice, with each location contains $q$ shells of one type. $q$ follows a distribution $P(q) \sim q^{-\alpha}$, which is motivated by the heterogeneity in a research subject's potential to yield papers: while some are fruitful, many of them are not. Hence, the seashore lattice contains multiple piles of shells, separated by a random number of empty sites in-between. A scientist starts at a random initial position and performs an unbiased random walk on its own lattice (50\% chance to move one step left or right). Upon reaching a site that contains shells, the walker picks one shell, corresponding to publishing one paper. The walker stops when it reaches the end of the career, defined by the total number of steps $S$ following a truncated log-normal distribution $P(S)$ motivated by the distribution of real career lifetime span observed \cite{petersen2012persistence, petersen2014reputation} (Supplementary Fig. S2).

Despite its simplicity, the ``seashore walk'' predicts the presence of both heterogeneity and recency features. If we assume that one type of shells corresponds to one research subject, the power-law distribution $P(q)$ provides varying limits to the number of papers that can be published on different research subjects. The unbiased random walker is likely to return to a site repeatedly, which consequently enables it to collect all shells on the site. Such repeated visits and $P(q)$ give rise to the heterogeneity and recency features observed empirically (Supplementary Figs. S3a, b).

To further capture the evolution of research interest, we need to assign topics to each type of shells. Recent advances in knowledge expansion \cite{rzhetsky2015choosing,foster2013tradition,cokol2005emergent} have provided two specific features governing the evolution of a scientist's research agenda: first, existing topics are connected to form a research subject; and second, new topics are occasionally added \cite{cokol2005emergent}. By absorbing these two features into the seashore walk, we define how the research subject for a type of shell is associated with its location on the lattice and how topics evolve on the underlying lattice, leading to the introduction of Model I (Fig. 3a).\\
Model I: We assume that there is a topic pool containing 3 topics underlying a site of lattice. The shells on the site, if there are any, are characterized by a random combination of the 3 topics (with repetitions), representing an artificial topic tuple for its research subject. The value 3 is based on the observation that each paper is characterized by 2.89 PACS codes on average \cite{radicchi2009diffusion} (Supplementary Figs. S4a, b). Furthermore, we assume that one topic pool covers $L$ sites on the lattice and two neighboring topic pools differ by one topic. Hence new topics are encountered as the walker moves away from the starting point, which is in line with the empirical observation that new topics emerge when the number of distinct topic tuples used by a scientist increases (Supplementary Fig. S4c).

Model I generates a sequence of shells for each walker traversing its own seashore, and each shell is characterized by an artificial topic tuple. By measuring interest change for an ensemble of walkers (see Supplementary Note 8 for the implementation of the simulation), we obtain $P(J)$ similar to that in real data (Fig. 4a and see Supplementary Note 9 for statistical analyses). Despite its simplicity, Model I fairly captures the process of research interest evolution. It also arises an interesting question: To what extent could the framework of seashore walk be improved? Note that the generation of topic tuples in Model I is based on simple processes, which can limit its capability to accurately reproduce the interest change distribution. If this is the case, we would expect a better result if more complexities are added to ensure that a walker's topic tuples as well as the correlations among these topic tuples are statistically similar to data. To test this conjecture, we construct Model II (Fig. 3b).\\
Model II: We generate a sequence of shells picked by a walker and identify the number of distinct shells in the sequence. For a walker who has picked $x$ distinct shells, we randomly find a scientist from the data who used $x$ different topic tuples in the publication sequence and randomly map them to the $x$ distinct shells (see Supplementary Note 10 for the implementation of simulation).

We apply Model II to the same shell sequences generated in Model I and find that the resulting research interest change distribution matches closer with the empirical observation (Fig. 4b and Supplementary Note 9 for statistical analyses). The improved result affirms the validity and potential of the seashore walk to capture the research interest evolution. It is noteworthy that the random mapping procedure in Model II is used only to avoid introducing any sophisticated means to generate topic tuples. It is inherently different from duplicating the actual sequence (Supplementary Discussion 1). Moreover, despite some assumptions defining the seashore walk, such as assuming the number of shells at a site follows a power-law distribution, the systematic reproduction of empirical observations arises from the interplay of multiple mechanisms. Missing either of them could invalidate the model (Supplementary Discussion 2).

Finally, the seashore walk makes two additional predictions regarding an individual's career. First, the individual's publication process is bursty, as the inter-publication time follows a power-law distribution \cite{barabasi2005origin,rybski2009scaling} (Supplementary Fig. S5a). In the model, a random walker's first passage steps asymptotically follow a power-law distribution with exponential cutoff $\sim S^{-3/2}$ \cite{redner2001guide}, giving rise to the burstiness of publication time (Fig. 4c). Second, the number of papers authored by a scientist follows a power-law distribution with exponential cutoff \cite{newman2001structure, araujo2014collaboration} (Supplementary Fig. S5b). We obtain the same form of distribution from the model (Fig. 4c). This is due to a combination of factors including the uniform probability of encountering sites with shells, the property that the mean number of sites visited by a random walker scales as $\sim S^{-1/2}$, and the existence of fat-tail in the log-normal distribution $P(S)$ (Supplementary Discussion 3).

The success of our simple model in capturing patterns observed in individual careers raises another question: Can other related approaches be adopted and applied to modeling the process investigated here? To this end, we identify two classes of models that might have been considered suitable based on existing works in science of science and network science. 
The first class pertains to models for individuals' mobility patterns \cite{song2010modelling, zhao2013emergence} by treating topic tuples as locations. In such models, a scientist's sequence of research subjects becomes the sequence of locations visited in an individual mobility trajectory (see Methods). While this approach would reproduce the heterogeneity feature based on preferential attachment \cite{Barabasi-Science-99}, it would not capture the recency feature. Indeed, under the preferential attachment mechanism, a positive feedback would arise by which the more frequently a research subject were studied, the more likely it would be studied again. As a result, an old subject would receive more attention than the recent one. Therefore, the probability of reusing a topic tuple ($\Pi$) would decrease with the rank of this tuple's first usage \cite{song2010modelling}. This, however, directly contradicts the recency feature observed in the data (Fig. 2e), demonstrating the inherent inability of preferential attachment based models to capture research interest change.
The second class of models treats the individual interest change as a Markov process on the knowledge network \cite{rzhetsky2015choosing, foster2013tradition, shi2015weaving}. This approach provides a comprehensive picture of the geometry of knowledge network that gives rise to subject proximity. However, the heterogeneity feature leading to the power-law distributed topic tuple usage can not be generated by a Markov process with fixed transition probabilities. Moreover, the knowledge network characterized by individuals' move between research subjects is not static but dynamic \cite{scholtes2014causality}, which can not be accounted for without introducing a much more complicated model.
Taken together, both approaches exhibit clear limitations in reproducing important characteristics of interest evolution studied in this paper. Our model, to the contrary, overcomes these limitations and preserves the patterns observed in the research interest evolution.

In summary, by taking advantage of the PACS codes that classify general areas of physics into multiple clearly-defined sub-areas, we quantitatively measure the extent of interest change for over 14,000 scientists, and discovered an exponential distribution of interest change within the population. We identify three key features in interest evolution essential for the presence of observed distribution. We further develop a simple statistical model that describes scientific research as a random walk and successfully captures empirical observations. Together, our results fill a critical gap in our quantitative understanding of science at large scale by identifying a set of macroscopic patterns governing research interest change throughout individual careers. Despite the well-known fact that scientists' choices of research subjects are driven by myriad of factors, our results indicate that research interest evolution can be well captured by a simple statistical model, uncovering a new degree of regularity underlying individual careers.

The methodology introduced here implies some limitations and potential for future work. When composing topic vector, we assume papers on which an author's name appears are equally representative of his/her research interests. This assumption is justified by the difference between the interest in the problem addressed by this paper and the contribution to the paper or recognition of each author \cite{merton1973sociology, shen2014collective}: every author has to be interested in the problem to engage in co-authoring the paper. In the future work, it would be interesting to systematically quantify the difference of each co-author's interest in a single paper on which they collaborate. The macroscopic patterns emphasized in this study are not significantly affected by potential errors in name disambiguation given the large number of scientists analyzed (Supplementary Note 1). Yet, the accuracy of author name disambiguation needs to be constantly challenged and scrutinized whenever publication data is applied. The systematic nature of classification codes and their rich, hierarchical structures make them good approximations of topics in research ranging from scientific discoveries \cite{herrera2010mapping, radicchi2011rescaling, pan2012evolution, wei2013scientists, sinatra2015century, shen2016interrelations} to inventions \cite{strumsky2012using, youn2015invention}. But we need to understand better the degree to which classification codes are good proxies for research topics. Our ability to identify author names and research topics of papers may improve dramatically, however, thanks to rapid advances in AI and natural language processing (NLP) that may offer more comprehensive publication data sets in the near future.

Promising future directions include extending our simple model proposed here to a multidimensional random walk in the knowledge space, which may lead to a model capturing a richer set of phenomena characterizing individual careers. Other directions include extending this work to other scientific domains to address the universality and robustness of our results, 
investigating how the observed patterns depend on contextual information such as nations \cite{zhang2013characterizing, king2004scientific}, institutions \cite{deville2014career, clauset2015systematic}, scientific disciplines \cite{hoonlor2013trends}, size of the research community \cite{jones2008multi, milojevic2014principles,vilhena2014finding}, status of a scientist \cite{merton1973sociology, petersen2012persistence, petersen2014reputation}, and publication habits. 
It would also be important to understand the short-term benefit and long-term scientific impact \cite{bergstrom2008eigenfactor, radicchi2008universality, wang2013quantifying, yao2014ranking, ke2015defining, sinatra2016quantifying} of research interest change by focusing on citations instead of publications. 
Answering these questions could not only offer a better understanding of the fundamental mechanisms underpinning a scientific career, but might also substantially improve our ability to trace, assess, predict and nurture high-impact scientists.

{\bf Methods}

{\bf Calculating topic vector.} The value of each element in the topic vector represents a topic's normalized frequency of occurrence in the set of papers analyzed. Given a topic tuple, we can use a vector $X = (a_1, a_2,\ldots,a_{67})$ to express each of the 67 topics' occurrence in the topic tuple, where $a_i = 0$ means the $i^\text{th}$ topic is not included in the topic tuple, $a_i = 1$ means that the $i^\text{th}$ topic appears once, and so on. $N_X = \sum_{i=1}^{67}a_i$ is the size of the topic tuple. By normalizing $X$, we obtain a vector $Y = (b_1, b_2,\ldots,b_{67})$ in which $b_i = a_i/N_X$ is the normalized frequency of occurrence of the $i^\text{th}$ topic. The topic vector $g$ is calculated by averaging $m$ different $Y$ vectors drawn from $m$ papers, as $g = \sum_{j=1}^{m}Y_j/m$. Take the calculation of $g_i$ in Fig. 1 as an example. The two topic tuples for $g_i$ are $(68,89,89)$ and $(02,05,68)$. The element value of topic 68 is calculated as $\frac{1/3 + 1/3}{2} = \frac{1}{3}$ as it appears once in each of the topic tuple. The element value of topic 89 is calculated as $\frac{2/3+0}{2} = \frac{1}{3}$ as it appears twice in one topic tuple and is not included in the other. The elemental values of topic 02 and 05 are calculated as $\frac{1/3+0}{2} = \frac{1}{6}$.

{\bf Measuring $\Delta n$.} $\Delta n$ is measured as the separation between the appearances of the same topic tuple in a scientist's publication sequence. For example, representing distinct topic tuples as different capital letters and assuming a publication sequence is ``A A A B C C B D E'', we obtain the following series of measurements: $\Delta n_\text{A} = 1$, $\Delta n_\text{A} = 1$, $\Delta n_\text{B} = 3$ and $\Delta n_\text{C} = 1$. These values are then used to calculate the distribution $P(\Delta n)$.

{\bf Measuring $\Pi$.} Each publication sequence is characterized by two parameters. One is the number of papers $n$ (i.e.\ the length of the sequence) and the other is the number of distinct topic tuples in the sequence ($x$). As both parameters vary among individuals, we first fix a set of distinct topic tuples to measure their re-usage frequency $\Pi$. Here we focus on the first 5 distinct topic tuples in the sequence (i.e.\ maximum rank is 5). Therefore, only those sequences with $x \ge 5$ are considered. We also analyze other cases by filtering $x \ge 4$ and $x \ge 6$ and similar patterns are observed. For each qualified sequence ($x \ge 5$), we go through it from the beginning until the fifth distinct topic tuple is firstly used. We then start to count the instance when one of the five topic tuples is re-used. For each individual, we obtain a fraction of time each of the five topic tuples is re-used. Such the fraction is then averaged over all qualified sequences to generate $\Pi$.

{\bf Generating the sequence using a preferential attachment based model.} We apply a preferential attachment based model to generated topic tuple sequences with power-law distributed usage of each topic tuple \cite{Barabasi-Science-99, song2010modelling, zhao2013emergence}. In the model, an individual's activity is randomly chosen from the two actions. One is to explore a new subject and publish a paper with a new topic tuple. The other is to return to a previously studied subject and publish a paper with a topic tuple already used. The probability to explore is defined as $\rho n^{-\gamma}$ in which the term $n^{-\gamma}$ captures the decreasing trend to explore a new subject as the number of papers increases. Consequently the probability of return, i.e.\ to reuse an old topic tuple, is $1 - \rho n^{-\gamma}$. If one returns, the choice of existing topic tuples is governed by preferential attachment: the probability $p_i$ to use a specific topic tuple $i$ is proportional to the tuple $i$'s current usage, so $p_i = n_i/\sum_{j}{n_j}$ where $n_i$ is the number of times that the tuple $i$ is used. The parameters applied are $\rho = 0.4$ and $\gamma = 0.1$. Each individual's time step is controlled by the number of papers published, following distribution $P(n) \sim n^{-1.5}$ with a cutoff $n_\text{max} = 150$. These variables make the sequence generated similar to those in real data. We generate totally 20,000 independent sequences, a comparable number to the size of real data. See Supplementary Discussion 4 for more about the preferential attachment based model.

{\bf Data availability.} The Physical Review dataset is available upon requested from the APS at\\ http://journals.aps.org/datasets. The name disambiguation procedure and the associated data are described in Ref.\cite{deville2014career,sinatra2016quantifying}.

{\bf Code availability.} Computational codes for data processing, analysis, and model simulation are available upon request.

{\bf Acknowledgments}
We thank Albert-Laszlo Barabasi for generously providing the initial dataset. We thank A.-L. Barab\'{a}si and G. Korniss for helpful discussions. This work is supported by the Army Research Laboratory under Cooperative Agreement Number W911NF-09-2-0053. T.J. is supported by the Natural Science Foundation of China (No. 6160309) and CCF-Tencent RAGR (No. 20160107). D.W. is supported by the Air Force Office of Scientific Research under award number FA9550-15-1-0162 and FA9550-17-1-0089. The funders had no role in study design, data collection and analysis, decision to publish, or preparation of the manuscript.

{\bf Author contributions}
 T.J., D.W and B.K.S. designed the research. T.J. performed numerical simulation and analyzed the empirical data. T.J., D.W. and B.K.S. prepared the paper.

{\bf Correspondence and requests for materials} should be addressed to  T.J. or D.W. or B.K.S.

{\bf Competing interests:} The authors declare no competing interests.

\newpage
\begin{figure}[h]
\begin{center}
\resizebox{9cm}{!}{\includegraphics{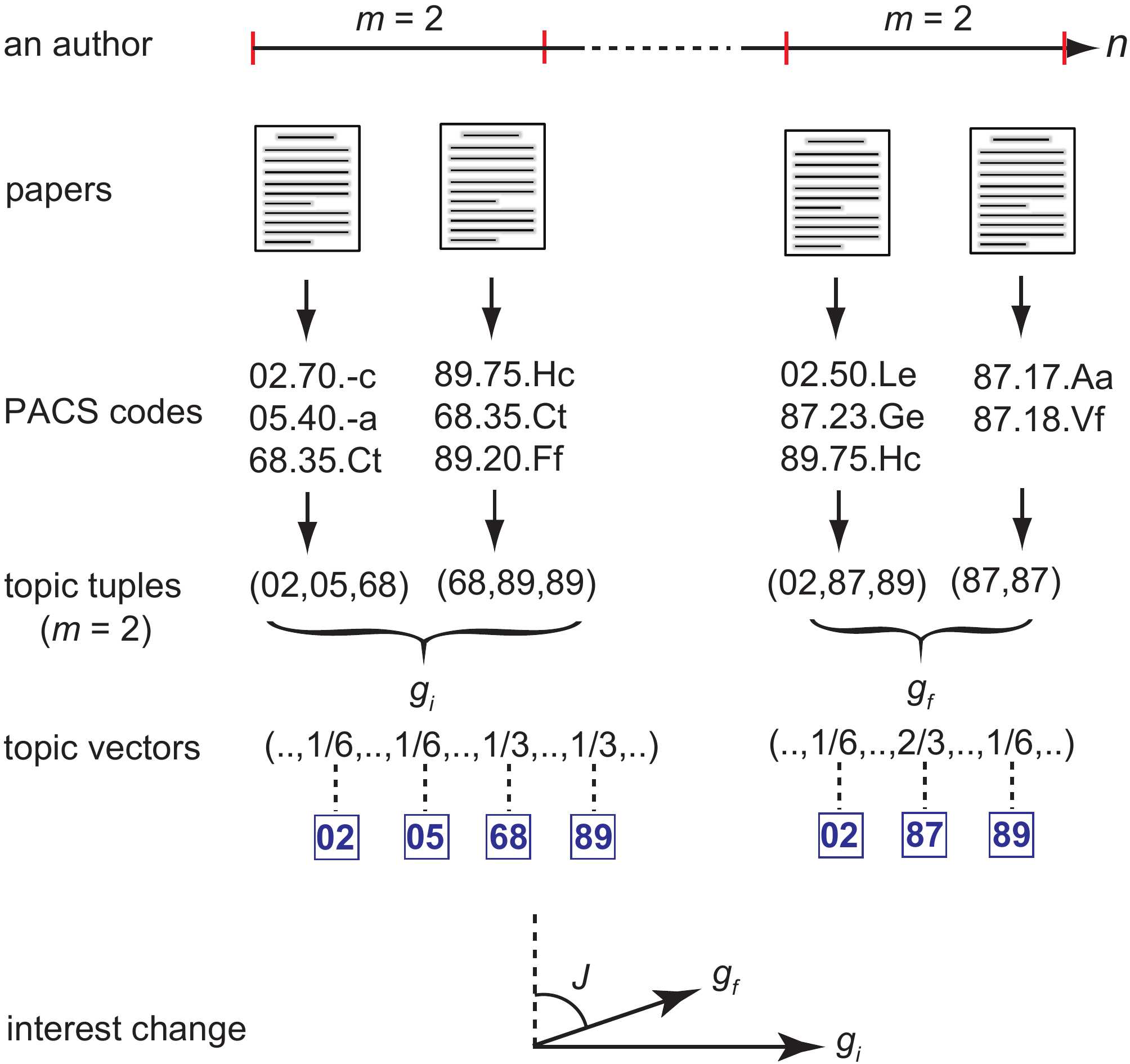}}
\caption{{\bf An example demonstrating the procedure to compose topic tuple and topic vector.} Using two topic vectors $g_i$ and $g_f$ based on the first and last $m$ papers ($m=2$ in this case) in an author's publication sequence, the interest change $J$ is calculated as the complementary cosine similarity between the two vectors. More details about calculating topic vector are given in Methods.
\label{fig:fig1}}
\end{center}
\end{figure}\noindent

\newpage

\begin{figure}[h]
\begin{center}
\resizebox{12cm}{!}{\includegraphics{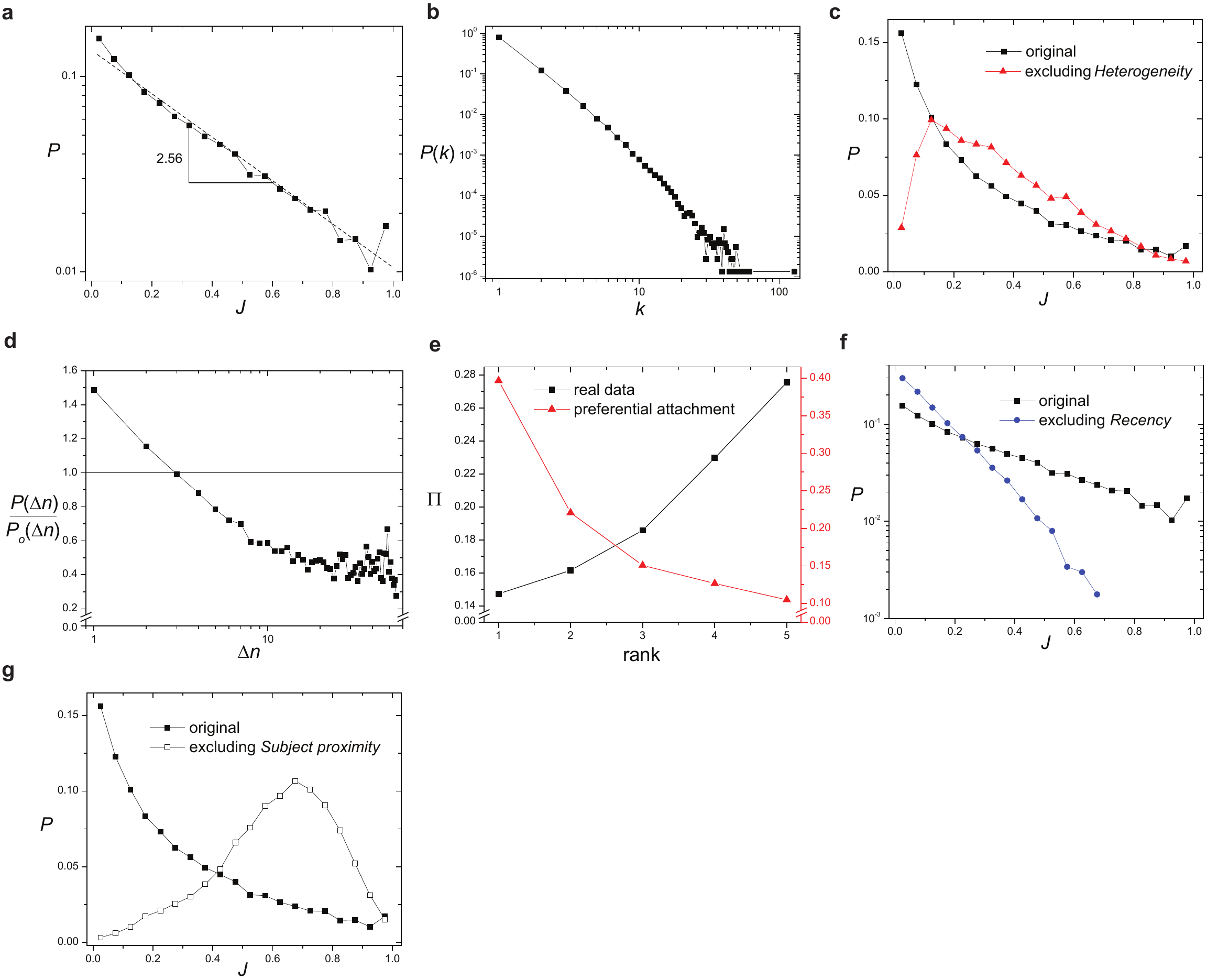}}
\caption{{\bf Patterns in the research interest evolution.}
{\bf a}, The fraction of scientists $P$ within a given range of interest change $(J-0.025,J+0.025]$ drops monotonically with $J$, which can be well fitted with an exponential function. At the boundary $J=0$, the range $[0, 0.05]$ is used, and the same boundary condition applies in all the following studies.
{\bf b}, The probability $P(k)$ that a topic tuple is used for $k$ times in one's publication sequence follows a power-law distribution, documenting the heterogeneity in the topic tuple usage.
{\bf c}, $P(J)$ over sequences in which {\it Heterogeneity} is eliminated by retaining only the first occurrence of each topic tuple in the publication sequence.
{\bf d}, The separation between the usage of the same topic tuple in a scientist's publication sequence, measured by the number of papers $\Delta n$ between its two consecutive uses (see Methods). The ratio between the distribution of $\Delta n$ of real data ($P(\Delta n)$) and that of the reshuffled sequence ($P_\text{o}(\Delta n)$) implies that an author is more likely to publish on subjects covered in recent papers than on those published long time ago.
{\bf e}, The relationship between the probability to reuse a previously studied topic tuple $\Pi$ (see Methods) and the rank of its first usage (rank 1 is assigned to the first distinct topic tuple used in an individual's career, $etc$). It demonstrates that a scientist more likely to publish on subjects recently studied than ones investigated long time ago. This is, however, opposite to what preferential attachment would predict (see Methods).
{\bf f}, $P(J)$ over sequences in which {\it Recency} is eliminated by removing the temporal correlation of topics through random shuffling of papers. 
{\bf g}, $P(J)$ over sequences in which {\it Subject proximity} is eliminated by replacing the distinct topic tuples in each publication sequence with ones randomly drawn from all existing topic tuples.
In all ({\bf c}), ({\bf f}) and ({\bf g}), p-value equals 0 in the two-sample Kolmogorov-Smirnov test, indicating that the observed differences are statistically significant.
\label{fig:data}}
\end{center}
\end{figure}\noindent 

\newpage

\begin{figure}[h]
\begin{center}
\resizebox{12cm}{!}{\includegraphics{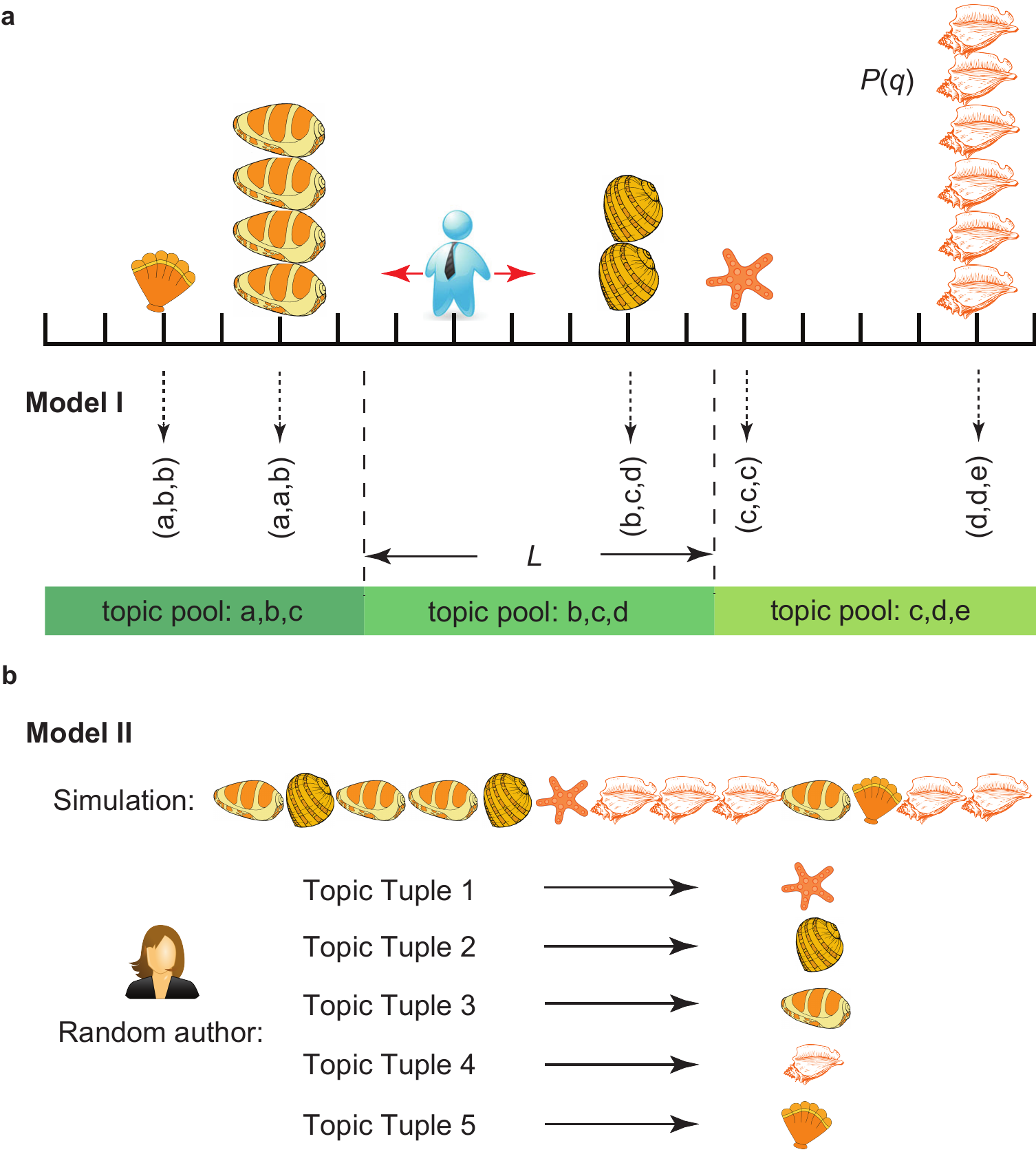}}
\caption{{\bf An illustration of the ``seashore walk''.} {\bf a}, The random walker traverses a 1-D lattice with piles of ``shells'' located at sites of the lattice. The probability that a site contains any shells is $p$. The number of shells at a non-empty site is characterized by the distribution $P(q)$. The walker picks a shell upon reaching a site that contains shells, corresponding to publishing a paper. A sequence of shells is generated till the walker stops after exhausting the total number of steps $S$ assigned to its career span, which is characterized by a log-normal distribution $P(S)$. In Model I, we assign each shell an artificial topic tuple based on the location at which the shell is picked. In particular, $L$ sites on the lattice share a topic pool with 3 topics. The two neighboring topic pools vary from one to the other by exactly one topic. For example, if one topic pool is ``a, b, c" then the next pool could be ``b, c, d", and so on, where the codes ``a, b, $\ldots$" represent any arbitrary characterizations of different topics. The ``shells'' at a site are characterized by an artificial topic tuple as a set of 3 topics, each randomly drawn from the topic pool below.
{\bf b}, In Model II, distinct shells picked by a walker are assigned to real topic tuples used by a random scientist. 
\label{fig:model}
}
\end{center}
\end{figure}\noindent

\newpage

\begin{figure}[h]
\begin{center}
\resizebox{12cm}{!}{\includegraphics{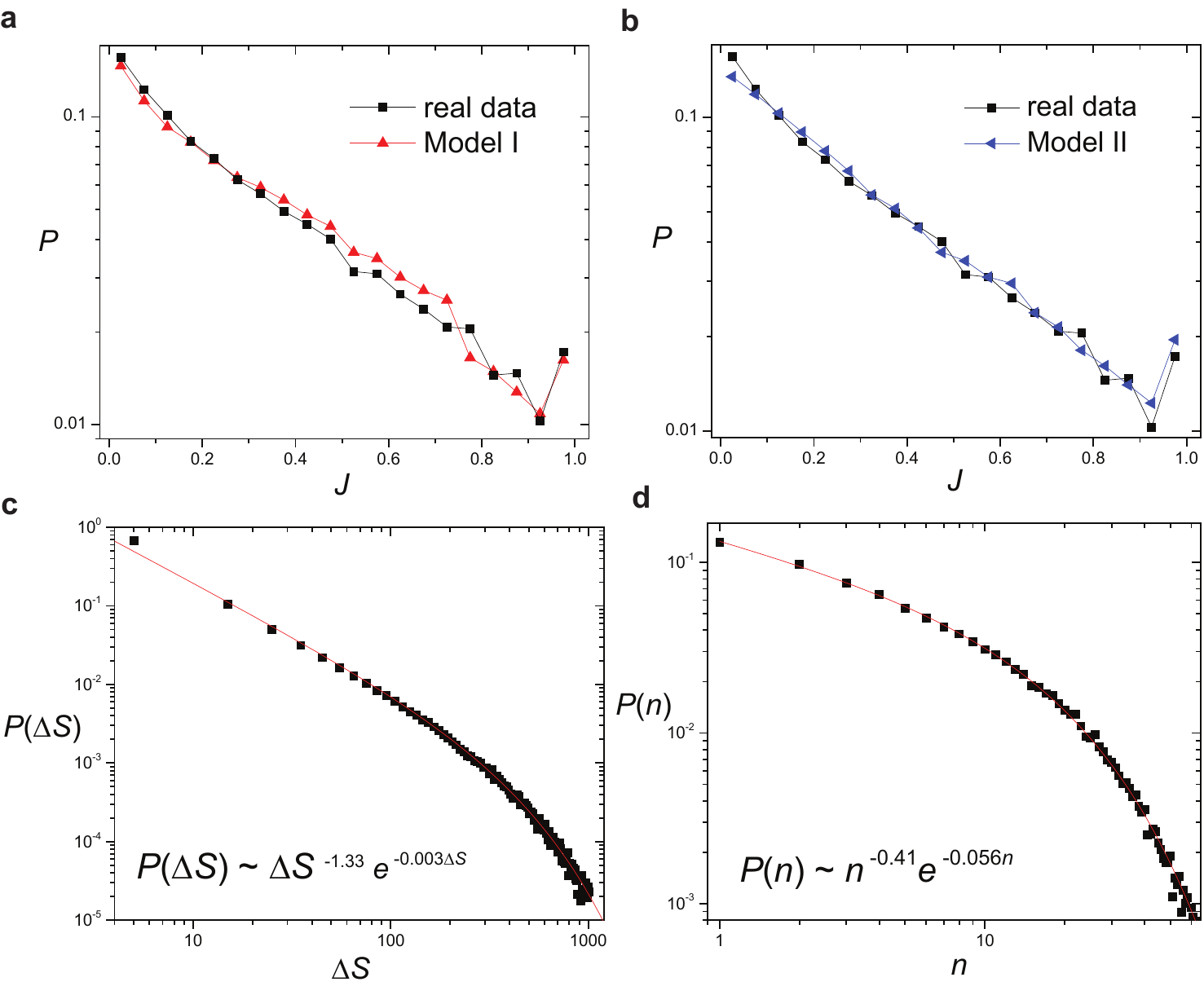}}
\caption{{\bf Results of the ``seashore walk''.}
{\bf a}, Model I generates interest change distribution qualitatively similar to that in real data. The variables are $p=0.2$, $L=35$, $P(q) \sim q^{-2}$, and the log-normal distribution $P(S)$ with mean $\mu=6$, standard deviation $\sigma = 3$ and cutoff $S_\text{max} = 2,000$.
{\bf b}, For the same sequence of shells in {\bf a}, Model II produces interest change distribution that matches closely with that in real data.
{\bf c}, The distribution of intervals between a scientist's successive publications $P(\Delta S)$ is characterized by a power-law distribution with exponential cutoff in our model.
{\bf d}, The number of papers authored by a scientist follows a power-law distribution with exponential cutoff in our model. The survival functions of {\bf c} and {\bf d} are plotted in Supplementary Fig. S6.
\label{fig:model_result}
}
\end{center}
\end{figure}\noindent


\begin{thebibliography}{10}
\expandafter\ifx\csname url\endcsname\relax
  \def\url#1{\texttt{#1}}\fi
\expandafter\ifx\csname urlprefix\endcsname\relax\def\urlprefix{URL }\fi
\providecommand{\bibinfo}[2]{#2}
\providecommand{\eprint}[2][]{\url{#2}}

\bibitem{kuhn2012structure}
\bibinfo{author}{Kuhn, T.~S.}
\newblock \emph{\bibinfo{title}{The structure of scientific revolutions}}
  (\bibinfo{publisher}{University of Chicago press}, \bibinfo{year}{2012}).

\bibitem{de1986little}
\bibinfo{author}{de~Solla~Price, D.~J.}
\newblock \emph{\bibinfo{title}{Little science, big science... and beyond}}
  (\bibinfo{publisher}{Columbia University Press New York},
  \bibinfo{year}{1986}).

\bibitem{merton1973sociology}
\bibinfo{author}{Merton, R.~K.}
\newblock \emph{\bibinfo{title}{The sociology of science: Theoretical and
  empirical investigations}} (\bibinfo{publisher}{University of Chicago press},
  \bibinfo{year}{1973}).

\bibitem{rzhetsky2015choosing}
\bibinfo{author}{Rzhetsky, A.}, \bibinfo{author}{Foster, J.~G.},
  \bibinfo{author}{Foster, I.~T.} \& \bibinfo{author}{Evans, J.~A.}
\newblock \bibinfo{title}{Choosing experiments to accelerate collective
  discovery}.
\newblock \emph{\bibinfo{journal}{Proc. Natl. Acad. Sci.}}
  \textbf{\bibinfo{volume}{112}}, \bibinfo{pages}{14569--14574}
  (\bibinfo{year}{2015}).

\bibitem{kuhn1979essential}
\bibinfo{author}{Kuhn, T.~S.}
\newblock \bibinfo{title}{The essential tension}.
\newblock \emph{\bibinfo{journal}{Selected Studies in Scientific Tradition and
  Change}}  (\bibinfo{year}{1979}).

\bibitem{sinatra2016quantifying}
\bibinfo{author}{Sinatra, R.}, \bibinfo{author}{Wang, D.},
  \bibinfo{author}{Deville, P.}, \bibinfo{author}{Song, C.} \&
  \bibinfo{author}{Barab{\'a}si, A.-L.}
\newblock \bibinfo{title}{Quantifying the evolution of individual scientific
  impact}.
\newblock \emph{\bibinfo{journal}{Science}} \textbf{\bibinfo{volume}{354}},
  \bibinfo{pages}{aaf5239} (\bibinfo{year}{2016}).

\bibitem{wang2013quantifying}
\bibinfo{author}{Wang, D.}, \bibinfo{author}{Song, C.} \&
  \bibinfo{author}{Barab{\'a}si, A.-L.}
\newblock \bibinfo{title}{Quantifying long-term scientific impact}.
\newblock \emph{\bibinfo{journal}{Science}} \textbf{\bibinfo{volume}{342}},
  \bibinfo{pages}{127--132} (\bibinfo{year}{2013}).

\bibitem{foster2013tradition}
\bibinfo{author}{Foster, J.~G.}, \bibinfo{author}{Rzhetsky, A.} \&
  \bibinfo{author}{Evans, J.~A.}
\newblock \bibinfo{title}{Tradition and innovation in scientists' research
  strategies}.
\newblock \emph{\bibinfo{journal}{Am. Sociol. Rev.}}
  \textbf{\bibinfo{volume}{80}}, \bibinfo{pages}{875--908}
  (\bibinfo{year}{2015}).

\bibitem{clauset2015systematic}
\bibinfo{author}{Clauset, A.}, \bibinfo{author}{Arbesman, S.} \&
  \bibinfo{author}{Larremore, D.~B.}
\newblock \bibinfo{title}{Systematic inequality and hierarchy in faculty hiring
  networks}.
\newblock \emph{\bibinfo{journal}{Sci. Adv.}} \textbf{\bibinfo{volume}{1}},
  \bibinfo{pages}{e1400005} (\bibinfo{year}{2015}).

\bibitem{jones2011age}
\bibinfo{author}{Jones, B.~F.} \& \bibinfo{author}{Weinberg, B.~A.}
\newblock \bibinfo{title}{Age dynamics in scientific creativity}.
\newblock \emph{\bibinfo{journal}{Proc. Natl. Acad. Sci.}}
  \textbf{\bibinfo{volume}{108}}, \bibinfo{pages}{18910--18914}
  (\bibinfo{year}{2011}).

\bibitem{packalen2015multiple}
\bibinfo{author}{Packalen, K.}
\newblock \bibinfo{title}{Multiple successful models: how demographic features
  of founding teams differ between regions and over time}.
\newblock \emph{\bibinfo{journal}{Entrepreneurship \& Regional Development}}
  \bibinfo{pages}{1--29} (\bibinfo{year}{2015}).

\bibitem{duch2012possible}
\bibinfo{author}{Duch, J.} \emph{et~al.}
\newblock \bibinfo{title}{The possible role of resource requirements and
  academic career-choice risk on gender differences in publication rate and
  impact.}
\newblock \emph{\bibinfo{journal}{PLoS ONE}} \textbf{\bibinfo{volume}{7}}
  (\bibinfo{year}{2012}).

\bibitem{west2013role}
\bibinfo{author}{West, J.~D.}, \bibinfo{author}{Jacquet, J.},
  \bibinfo{author}{King, M.~M.}, \bibinfo{author}{Correll, S.~J.} \&
  \bibinfo{author}{Bergstrom, C.~T.}
\newblock \bibinfo{title}{The role of gender in scholarly authorship}.
\newblock \emph{\bibinfo{journal}{PloS one}} \textbf{\bibinfo{volume}{8}},
  \bibinfo{pages}{e66212} (\bibinfo{year}{2013}).

\bibitem{malmgren2010role}
\bibinfo{author}{Malmgren, R.~D.}, \bibinfo{author}{Ottino, J.~M.} \&
  \bibinfo{author}{Amaral, L. A.~N.}
\newblock \bibinfo{title}{The role of mentorship in prot{\'e}g{\'e}
  performance}.
\newblock \emph{\bibinfo{journal}{Nature}} \textbf{\bibinfo{volume}{465}},
  \bibinfo{pages}{622--626} (\bibinfo{year}{2010}).

\bibitem{guimera2005team}
\bibinfo{author}{Guimera, R.}, \bibinfo{author}{Uzzi, B.},
  \bibinfo{author}{Spiro, J.} \& \bibinfo{author}{Amaral, L. A.~N.}
\newblock \bibinfo{title}{Team assembly mechanisms determine collaboration
  network structure and team performance}.
\newblock \emph{\bibinfo{journal}{Science}} \textbf{\bibinfo{volume}{308}},
  \bibinfo{pages}{697--702} (\bibinfo{year}{2005}).

\bibitem{jones2008multi}
\bibinfo{author}{Jones, B.~F.}, \bibinfo{author}{Wuchty, S.} \&
  \bibinfo{author}{Uzzi, B.}
\newblock \bibinfo{title}{Multi-university research teams: shifting impact,
  geography, and stratification in science}.
\newblock \emph{\bibinfo{journal}{Science}} \textbf{\bibinfo{volume}{322}},
  \bibinfo{pages}{1259--1262} (\bibinfo{year}{2008}).

\bibitem{hoonlor2013trends}
\bibinfo{author}{Hoonlor, A.}, \bibinfo{author}{Szymanski, B.~K.} \&
  \bibinfo{author}{Zaki, M.~J.}
\newblock \bibinfo{title}{Trends in computer science research}.
\newblock \emph{\bibinfo{journal}{Commun. ACM}} \textbf{\bibinfo{volume}{56}},
  \bibinfo{pages}{74--83} (\bibinfo{year}{2013}).

\bibitem{mcnally2011discovery}
\bibinfo{author}{McNally, A.}, \bibinfo{author}{Prier, C.~K.} \&
  \bibinfo{author}{MacMillan, D.~W.}
\newblock \bibinfo{title}{Discovery of an $\alpha$-amino c--h arylation
  reaction using the strategy of accelerated serendipity}.
\newblock \emph{\bibinfo{journal}{Science}} \textbf{\bibinfo{volume}{334}},
  \bibinfo{pages}{1114--1117} (\bibinfo{year}{2011}).

\bibitem{ericsson2006influence}
\bibinfo{author}{Ericsson, K.~A.}
\newblock \bibinfo{title}{The influence of experience and deliberate practice
  on the development of superior expert performance}.
\newblock \emph{\bibinfo{journal}{The Cambridge handbook of expertise and
  expert performance}} \bibinfo{pages}{683--703} (\bibinfo{year}{2006}).

\bibitem{azoulay2011incentives}
\bibinfo{author}{Azoulay, P.}, \bibinfo{author}{Graff~Zivin, J.~S.} \&
  \bibinfo{author}{Manso, G.}
\newblock \bibinfo{title}{Incentives and creativity: evidence from the academic
  life sciences}.
\newblock \emph{\bibinfo{journal}{RAND J Econ}} \textbf{\bibinfo{volume}{42}},
  \bibinfo{pages}{527--554} (\bibinfo{year}{2011}).

\bibitem{bergstrom2016scientists}
\bibinfo{author}{Bergstrom, C.~T.}, \bibinfo{author}{Foster, J.~G.} \&
  \bibinfo{author}{Song, Y.}
\newblock \bibinfo{title}{Why scientists chase big problems: Individual
  strategy and social optimality}.
\newblock \emph{\bibinfo{journal}{arXiv:1605.05822}}  (\bibinfo{year}{2016}).

\bibitem{march1991exploration}
\bibinfo{author}{March, J.~G.}
\newblock \bibinfo{title}{Exploration and exploitation in organizational
  learning}.
\newblock \emph{\bibinfo{journal}{Organ. Sci.}} \textbf{\bibinfo{volume}{2}},
  \bibinfo{pages}{71--87} (\bibinfo{year}{1991}).

\bibitem{vilhena2014finding}
\bibinfo{author}{Vilhena, D.~A.} \emph{et~al.}
\newblock \bibinfo{title}{Finding cultural holes: how structure and culture
  diverge in networks of scholarly communication}.
\newblock \emph{\bibinfo{journal}{Sociological Science}}
  \textbf{\bibinfo{volume}{1}}, \bibinfo{pages}{221--238}
  (\bibinfo{year}{2014}).

\bibitem{hidalgo2015information}
\bibinfo{author}{Hidalgo, C.}
\newblock \emph{\bibinfo{title}{Why Information Grows: The Evolution of Order,
  from Atoms to Economies}} (\bibinfo{publisher}{Basic Books},
  \bibinfo{year}{2015}).

\bibitem{de1966collaboration}
\bibinfo{author}{de~Solla~Price, D.~J.} \& \bibinfo{author}{Beaver, D.}
\newblock \bibinfo{title}{Collaboration in an invisible college.}
\newblock \emph{\bibinfo{journal}{Am. Psychol.}} \textbf{\bibinfo{volume}{21}},
  \bibinfo{pages}{1011} (\bibinfo{year}{1966}).

\bibitem{jones2009burden}
\bibinfo{author}{Jones, B.~F.}
\newblock \bibinfo{title}{The burden of knowledge and the “death of the
  renaissance man”: Is innovation getting harder?}
\newblock \emph{\bibinfo{journal}{Rev. Econ. Stud.}}
  \textbf{\bibinfo{volume}{76}}, \bibinfo{pages}{283--317}
  (\bibinfo{year}{2009}).

\bibitem{uzzi2013atypical}
\bibinfo{author}{Uzzi, B.}, \bibinfo{author}{Mukherjee, S.},
  \bibinfo{author}{Stringer, M.} \& \bibinfo{author}{Jones, B.}
\newblock \bibinfo{title}{Atypical combinations and scientific impact}.
\newblock \emph{\bibinfo{journal}{Science}} \textbf{\bibinfo{volume}{342}},
  \bibinfo{pages}{468--472} (\bibinfo{year}{2013}).

\bibitem{youn2015invention}
\bibinfo{author}{Youn, H.}, \bibinfo{author}{Strumsky, D.},
  \bibinfo{author}{Bettencourt, L.~M.} \& \bibinfo{author}{Lobo, J.}
\newblock \bibinfo{title}{Invention as a combinatorial process: evidence from
  us patents}.
\newblock \emph{\bibinfo{journal}{‎J. R. Soc. Interface}}
  \textbf{\bibinfo{volume}{12}}, \bibinfo{pages}{20150272}
  (\bibinfo{year}{2015}).

\bibitem{cokol2005emergent}
\bibinfo{author}{Cokol, M.}, \bibinfo{author}{Iossifov, I.},
  \bibinfo{author}{Weinreb, C.} \& \bibinfo{author}{Rzhetsky, A.}
\newblock \bibinfo{title}{Emergent behavior of growing knowledge about
  molecular interactions}.
\newblock \emph{\bibinfo{journal}{Nat. Biotechnol.}}
  \textbf{\bibinfo{volume}{23}}, \bibinfo{pages}{1243--1247}
  (\bibinfo{year}{2005}).

\bibitem{sinatra2015century}
\bibinfo{author}{Sinatra, R.}, \bibinfo{author}{Deville, P.},
  \bibinfo{author}{Szell, M.}, \bibinfo{author}{Wang, D.} \&
  \bibinfo{author}{Barab{\'a}si, A.-L.}
\newblock \bibinfo{title}{A century of physics}.
\newblock \emph{\bibinfo{journal}{Nat. Phys.}} \textbf{\bibinfo{volume}{11}},
  \bibinfo{pages}{791--796} (\bibinfo{year}{2015}).

\bibitem{gonzalez2008understanding}
\bibinfo{author}{Gonzalez, M.~C.}, \bibinfo{author}{Hidalgo, C.~A.} \&
  \bibinfo{author}{Barabasi, A.-L.}
\newblock \bibinfo{title}{Understanding individual human mobility patterns}.
\newblock \emph{\bibinfo{journal}{Nature}} \textbf{\bibinfo{volume}{453}},
  \bibinfo{pages}{779--782} (\bibinfo{year}{2008}).

\bibitem{song2010modelling}
\bibinfo{author}{Song, C.}, \bibinfo{author}{Koren, T.}, \bibinfo{author}{Wang,
  P.} \& \bibinfo{author}{Barab{\'a}si, A.-L.}
\newblock \bibinfo{title}{Modelling the scaling properties of human mobility}.
\newblock \emph{\bibinfo{journal}{Nat. Phys.}} \textbf{\bibinfo{volume}{6}},
  \bibinfo{pages}{818--823} (\bibinfo{year}{2010}).

\bibitem{simini2012universal}
\bibinfo{author}{Simini, F.}, \bibinfo{author}{Gonz{\'a}lez, M.~C.},
  \bibinfo{author}{Maritan, A.} \& \bibinfo{author}{Barab{\'a}si, A.-L.}
\newblock \bibinfo{title}{A universal model for mobility and migration
  patterns}.
\newblock \emph{\bibinfo{journal}{Nature}} \textbf{\bibinfo{volume}{484}},
  \bibinfo{pages}{96--100} (\bibinfo{year}{2012}).

\bibitem{yan2014universal}
\bibinfo{author}{Yan, X.-Y.}, \bibinfo{author}{Zhao, C.}, \bibinfo{author}{Fan,
  Y.}, \bibinfo{author}{Di, Z.} \& \bibinfo{author}{Wang, W.-X.}
\newblock \bibinfo{title}{Universal predictability of mobility patterns in
  cities}.
\newblock \emph{\bibinfo{journal}{‎J. R. Soc. Interface}}
  \textbf{\bibinfo{volume}{11}}, \bibinfo{pages}{20140834}
  (\bibinfo{year}{2014}).

\bibitem{barabasi2005origin}
\bibinfo{author}{Barabasi, A.-L.}
\newblock \bibinfo{title}{The origin of bursts and heavy tails in human
  dynamics}.
\newblock \emph{\bibinfo{journal}{Nature}} \textbf{\bibinfo{volume}{435}},
  \bibinfo{pages}{207--211} (\bibinfo{year}{2005}).

\bibitem{malmgren2008poissonian}
\bibinfo{author}{Malmgren, R.~D.}, \bibinfo{author}{Stouffer, D.~B.},
  \bibinfo{author}{Motter, A.~E.} \& \bibinfo{author}{Amaral, L.~A.}
\newblock \bibinfo{title}{A poissonian explanation for heavy tails in e-mail
  communication}.
\newblock \emph{\bibinfo{journal}{Proc. Natl. Acad. Sci.}}
  \textbf{\bibinfo{volume}{105}}, \bibinfo{pages}{18153--18158}
  (\bibinfo{year}{2008}).

\bibitem{zhao2013emergence}
\bibinfo{author}{Zhao, Z.-D.} \emph{et~al.}
\newblock \bibinfo{title}{Emergence of scaling in human-interest dynamics}.
\newblock \emph{\bibinfo{journal}{Sci. Rep.}} \textbf{\bibinfo{volume}{3}},
  \bibinfo{pages}{3472} (\bibinfo{year}{2013}).

\bibitem{Barabasi-Science-99}
\bibinfo{author}{Barab\'asi, A.-L.} \& \bibinfo{author}{Albert, R.}
\newblock \bibinfo{title}{Emergence of scaling in random networks}.
\newblock \emph{\bibinfo{journal}{Science}} \textbf{\bibinfo{volume}{286}},
  \bibinfo{pages}{509--512} (\bibinfo{year}{1999}).

\bibitem{scholtes2014causality}
\bibinfo{author}{Scholtes, I.} \emph{et~al.}
\newblock \bibinfo{title}{Causality-driven slow-down and speed-up of diffusion
  in non-markovian temporal networks}.
\newblock \emph{\bibinfo{journal}{Nat. Commun.}} \textbf{\bibinfo{volume}{5}},
  \bibinfo{pages}{5024} (\bibinfo{year}{2014}).

\bibitem{holme2012temporal}
\bibinfo{author}{Holme, P.} \& \bibinfo{author}{Saram{\"a}ki, J.}
\newblock \bibinfo{title}{Temporal networks}.
\newblock \emph{\bibinfo{journal}{Phys. Rep.}} \textbf{\bibinfo{volume}{519}},
  \bibinfo{pages}{97--125} (\bibinfo{year}{2012}).

\bibitem{zhang2013characterizing}
\bibinfo{author}{Zhang, Q.}, \bibinfo{author}{Perra, N.},
  \bibinfo{author}{Gon{\c{c}}alves, B.}, \bibinfo{author}{Ciulla, F.} \&
  \bibinfo{author}{Vespignani, A.}
\newblock \bibinfo{title}{Characterizing scientific production and consumption
  in physics}.
\newblock \emph{\bibinfo{journal}{Sci. Rep.}} \textbf{\bibinfo{volume}{3}},
  \bibinfo{pages}{1640} (\bibinfo{year}{2013}).

\bibitem{radicchi2009diffusion}
\bibinfo{author}{Radicchi, F.}, \bibinfo{author}{Fortunato, S.},
  \bibinfo{author}{Markines, B.} \& \bibinfo{author}{Vespignani, A.}
\newblock \bibinfo{title}{Diffusion of scientific credits and the ranking of
  scientists}.
\newblock \emph{\bibinfo{journal}{Phy. Rev. E}} \textbf{\bibinfo{volume}{80}},
  \bibinfo{pages}{056103} (\bibinfo{year}{2009}).

\bibitem{deville2014career}
\bibinfo{author}{Deville, P.} \emph{et~al.}
\newblock \bibinfo{title}{Career on the move: Geography, stratification, and
  scientific impact}.
\newblock \emph{\bibinfo{journal}{Sci. Rep.}} \textbf{\bibinfo{volume}{4}},
  \bibinfo{pages}{4770} (\bibinfo{year}{2014}).

\bibitem{herrera2010mapping}
\bibinfo{author}{Herrera, M.}, \bibinfo{author}{Roberts, D.~C.} \&
  \bibinfo{author}{Gulbahce, N.}
\newblock \bibinfo{title}{Mapping the evolution of scientific fields}.
\newblock \emph{\bibinfo{journal}{PloS one}} \textbf{\bibinfo{volume}{5}},
  \bibinfo{pages}{e10355} (\bibinfo{year}{2010}).

\bibitem{radicchi2011rescaling}
\bibinfo{author}{Radicchi, F.} \& \bibinfo{author}{Castellano, C.}
\newblock \bibinfo{title}{Rescaling citations of publications in physics}.
\newblock \emph{\bibinfo{journal}{Phys. Rev. E}} \textbf{\bibinfo{volume}{83}},
  \bibinfo{pages}{046116} (\bibinfo{year}{2011}).

\bibitem{pan2012evolution}
\bibinfo{author}{Pan, R.~K.}, \bibinfo{author}{Sinha, S.},
  \bibinfo{author}{Kaski, K.} \& \bibinfo{author}{Saram{\"a}ki, J.}
\newblock \bibinfo{title}{The evolution of interdisciplinarity in physics
  research}.
\newblock \emph{\bibinfo{journal}{Sci. Rep.}} \textbf{\bibinfo{volume}{2}},
  \bibinfo{pages}{551} (\bibinfo{year}{2012}).

\bibitem{wei2013scientists}
\bibinfo{author}{Wei, T.} \emph{et~al.}
\newblock \bibinfo{title}{Do scientists trace hot topics?}
\newblock \emph{\bibinfo{journal}{Sci. Rep.}} \textbf{\bibinfo{volume}{3}},
  \bibinfo{pages}{2207} (\bibinfo{year}{2013}).

\bibitem{shen2016interrelations}
\bibinfo{author}{Shen, Z.} \emph{et~al.}
\newblock \bibinfo{title}{Interrelations among scientific fields and their
  relative influences revealed by an input--output analysis}.
\newblock \emph{\bibinfo{journal}{J. Informetr.}}
  \textbf{\bibinfo{volume}{10}}, \bibinfo{pages}{82--97}
  (\bibinfo{year}{2016}).

\bibitem{shi2015weaving}
\bibinfo{author}{Shi, F.}, \bibinfo{author}{Foster, J.~G.} \&
  \bibinfo{author}{Evans, J.~A.}
\newblock \bibinfo{title}{Weaving the fabric of science: Dynamic network models
  of science's unfolding structure}.
\newblock \emph{\bibinfo{journal}{Social Networks}}
  \textbf{\bibinfo{volume}{43}}, \bibinfo{pages}{73--85}
  (\bibinfo{year}{2015}).

\bibitem{boyack2005mapping}
\bibinfo{author}{Boyack, K.~W.}, \bibinfo{author}{Klavans, R.} \&
  \bibinfo{author}{B{\"o}rner, K.}
\newblock \bibinfo{title}{Mapping the backbone of science}.
\newblock \emph{\bibinfo{journal}{Scientometrics}}
  \textbf{\bibinfo{volume}{64}}, \bibinfo{pages}{351--374}
  (\bibinfo{year}{2005}).

\bibitem{mandelbrote2001footprints}
\bibinfo{author}{Mandelbrote, S.}
\newblock \emph{\bibinfo{title}{Footprints of the Lion}}
  (\bibinfo{publisher}{Cambridge University Library}, \bibinfo{year}{2001}).

\bibitem{petersen2012persistence}
\bibinfo{author}{Petersen, A.~M.}, \bibinfo{author}{Riccaboni, M.},
  \bibinfo{author}{Stanley, H.~E.} \& \bibinfo{author}{Pammolli, F.}
\newblock \bibinfo{title}{Persistence and uncertainty in the academic career}.
\newblock \emph{\bibinfo{journal}{Proc. Natl. Acad. Sci.}}
  \textbf{\bibinfo{volume}{109}}, \bibinfo{pages}{5213--5218}
  (\bibinfo{year}{2012}).

\bibitem{petersen2014reputation}
\bibinfo{author}{Petersen, A.~M.} \emph{et~al.}
\newblock \bibinfo{title}{Reputation and impact in academic careers}.
\newblock \emph{\bibinfo{journal}{Proc. Natl. Acad. Sci.}}
  \textbf{\bibinfo{volume}{111}}, \bibinfo{pages}{15316--15321}
  (\bibinfo{year}{2014}).

\bibitem{rybski2009scaling}
\bibinfo{author}{Rybski, D.}, \bibinfo{author}{Buldyrev, S.~V.},
  \bibinfo{author}{Havlin, S.}, \bibinfo{author}{Liljeros, F.} \&
  \bibinfo{author}{Makse, H.~A.}
\newblock \bibinfo{title}{Scaling laws of human interaction activity}.
\newblock \emph{\bibinfo{journal}{Proc. Natl. Acad. Sci.}}
  \textbf{\bibinfo{volume}{106}}, \bibinfo{pages}{12640--12645}
  (\bibinfo{year}{2009}).

\bibitem{redner2001guide}
\bibinfo{author}{Redner, S.}
\newblock \emph{\bibinfo{title}{A guide to first-passage processes}}
  (\bibinfo{publisher}{Cambridge University Press}, \bibinfo{year}{2001}).

\bibitem{newman2001structure}
\bibinfo{author}{Newman, M.~E.}
\newblock \bibinfo{title}{The structure of scientific collaboration networks}.
\newblock \emph{\bibinfo{journal}{Proc. Natl. Acad. Sci.}}
  \textbf{\bibinfo{volume}{98}}, \bibinfo{pages}{404--409}
  (\bibinfo{year}{2001}).

\bibitem{araujo2014collaboration}
\bibinfo{author}{Ara{\'u}jo, E.~B.}, \bibinfo{author}{Moreira, A.~A.},
  \bibinfo{author}{Furtado, V.}, \bibinfo{author}{Pequeno, T.~H.} \&
  \bibinfo{author}{Andrade~Jr, J.~S.}
\newblock \bibinfo{title}{Collaboration networks from a large cv database:
  dynamics, topology and bonus impact}.
\newblock \emph{\bibinfo{journal}{PloS one}} \textbf{\bibinfo{volume}{9}},
  \bibinfo{pages}{e90537} (\bibinfo{year}{2014}).

\bibitem{shen2014collective}
\bibinfo{author}{Shen, H.-W.} \& \bibinfo{author}{Barab{\'a}si, A.-L.}
\newblock \bibinfo{title}{Collective credit allocation in science}.
\newblock \emph{\bibinfo{journal}{Proc. Natl. Acad. Sci.}}
  \textbf{\bibinfo{volume}{111}}, \bibinfo{pages}{12325--12330}
  (\bibinfo{year}{2014}).

\bibitem{strumsky2012using}
\bibinfo{author}{Strumsky, D.}, \bibinfo{author}{Lobo, J.} \&
  \bibinfo{author}{Van~der Leeuw, S.}
\newblock \bibinfo{title}{Using patent technology codes to study technological
  change}.
\newblock \emph{\bibinfo{journal}{Econ. Innov. N. Technol.}}
  \textbf{\bibinfo{volume}{21}}, \bibinfo{pages}{267--286}
  (\bibinfo{year}{2012}).

\bibitem{king2004scientific}
\bibinfo{author}{King, D.~A.}
\newblock \bibinfo{title}{The scientific impact of nations}.
\newblock \emph{\bibinfo{journal}{Nature}} \textbf{\bibinfo{volume}{430}},
  \bibinfo{pages}{311--316} (\bibinfo{year}{2004}).

\bibitem{milojevic2014principles}
\bibinfo{author}{Milojevi{\'c}, S.}
\newblock \bibinfo{title}{Principles of scientific research team formation and
  evolution}.
\newblock \emph{\bibinfo{journal}{Proc. Natl. Acad. Sci.}}
  \textbf{\bibinfo{volume}{111}}, \bibinfo{pages}{3984--3989}
  (\bibinfo{year}{2014}).

\bibitem{bergstrom2008eigenfactor}
\bibinfo{author}{Bergstrom, C.~T.}, \bibinfo{author}{West, J.~D.} \&
  \bibinfo{author}{Wiseman, M.~A.}
\newblock \bibinfo{title}{The eigenfactor metrics}.
\newblock \emph{\bibinfo{journal}{The Journal of Neuroscience}}
  \textbf{\bibinfo{volume}{28}}, \bibinfo{pages}{11433--11434}
  (\bibinfo{year}{2008}).

\bibitem{radicchi2008universality}
\bibinfo{author}{Radicchi, F.}, \bibinfo{author}{Fortunato, S.} \&
  \bibinfo{author}{Castellano, C.}
\newblock \bibinfo{title}{Universality of citation distributions: Toward an
  objective measure of scientific impact}.
\newblock \emph{\bibinfo{journal}{Proc. Natl. Acad. Sci.}}
  \textbf{\bibinfo{volume}{105}}, \bibinfo{pages}{17268--17272}
  (\bibinfo{year}{2008}).

\bibitem{yao2014ranking}
\bibinfo{author}{Yao, L.}, \bibinfo{author}{Wei, T.}, \bibinfo{author}{Zeng,
  A.}, \bibinfo{author}{Fan, Y.} \& \bibinfo{author}{Di, Z.}
\newblock \bibinfo{title}{Ranking scientific publications: the effect of
  nonlinearity}.
\newblock \emph{\bibinfo{journal}{Sci. Rep.}} \textbf{\bibinfo{volume}{4}},
  \bibinfo{pages}{6663} (\bibinfo{year}{2014}).

\bibitem{ke2015defining}
\bibinfo{author}{Ke, Q.}, \bibinfo{author}{Ferrara, E.},
  \bibinfo{author}{Radicchi, F.} \& \bibinfo{author}{Flammini, A.}
\newblock \bibinfo{title}{Defining and identifying sleeping beauties in
  science}.
\newblock \emph{\bibinfo{journal}{Proc. Natl. Acad. Sci.}}
  \textbf{\bibinfo{volume}{112}}, \bibinfo{pages}{7426--7431}
  (\bibinfo{year}{2015}).

\end{thebibliography}
\end{document}